\documentclass{osa-article}

\journal{osajournal}
\setprjcopyright

\articletype{Research Article}
\begin{document}

\title{Frequency multiplexed entanglement for continuous-variable quantum key distribution }

\author{Olena Kovalenko,\authormark{1,*} Young-Sik Ra,\authormark{2,3} Yin Cai,\authormark{2,4,5} Vladyslav C. Usenko,\authormark{1} Claude Fabre,\authormark{2}  Nicolas Treps\authormark{2} and Radim Filip\authormark{1}}
\address{\authormark{1}{Department of Optics, Palacky University, 17. listopadu 12, 77146 Olomouc, Czech Republic}\\
\authormark{2}{Laboratoire Kastler Brossel, Sorbonne Universit\'e, CNRS, ENS-Universit\'e PSL, Coll\`ege de France, 4 place Jussieu, 75252 Paris, France}\\
\authormark{3}{Department of Physics, Korea Advanced Institute of Science and Technology (KAIST), Daejeon 34141, Korea}\\
\authormark{4}{Key Laboratory for Physical Electronics and Devices of the Ministry of Education \& Shaanxi Key Lab of Information Photonic Technique, Xi\textsc{\char00}an Jiaotong University, 710049 Xi\textsc{\char00}an, China}}

\email{\authormark{*}kovalenko@optics.upol.cz}
\email{\authormark{5}caiyin@xjtu.edu.cn} %% email address is required

% \homepage{http:...} %% author's URL, if desired

%%%%%%%%%%%%%%%%%%% abstract %%%%%%%%%%%%%%%%
%% [use \begin{abstract*}...\end{abstract*} if exempt from copyright]

\begin{abstract}
Quantum key distribution with continuous variables already uses advantageous high-speed single-mode homodyne detection with low electronic noise at room temperature. Together with continuous-variable information encoding to nonclassical states, the distance for secure key transmission through lossy channels can approach 300 km in current optical fibers. Such protocols tolerate higher channel noise and also limited data processing efficiency compared to coherent-state protocols. The secret key  rate can be further increased by increasing the system clock rates, and, further, by a suitable frequency-mode-multiplexing of optical transmission channels. However, the multiplexed modes couple together in the source or any other part of the protocol. Therefore, multiplexed communication will experience crosstalk and the gain can be minuscule. Advantageously, homodyne detectors allow solving this crosstalk problem by proper data processing. It is a potential advantage over protocols with single-photon detectors, which do not enable similar data processing techniques. We demonstrate the positive outcome of this methodology on the experimentally characterized frequency-multiplexed entangled source of femtosecond optical pulses with natural crosstalk between eight entangled pairs of modes. As the main result, we predict almost 15-fold higher secret key rate. This experimental test and analysis of frequency-multiplexed entanglement source opens the way for the field implementation of  high-capacity quantum key distribution with continuous variables.
\end{abstract}

%%%%%%%%%%%%%%%%%%%%%%%%%%  body  %%%%%%%%%%%%%%%%%%%%%%%%%%

\section{Introduction} 

Quantum key distribution (QKD) \cite{Pirandola2019} is a pioneering application of quantum information theory enabled by fundamental particle and wave quantum features of light. Advantageously, in experiments at optical wavelengths, QKD can exploit complementary photon counting and homodyne detection methods of quantum optics. Naturally, both methods have advantages and disadvantages, fundamental as well as technical. Therefore, the optimal implementation of a quantum-secure network will be likely hybrid in the future, combining the advantages and suppressing the weaknesses of different protocols respectively to the requirements and conditions \cite{Cattaneo2018}. Currently, homodyne detection is fast, efficient and extremely low-noise, tolerant to background noise in the channel \cite{Chi2009}. This  hardware already opened space for a high-speed secret key generation. For a long time, the homodyne detection stimulated a large set of  theoretical proposals \cite{Ralph1999,Grosshans2002,Weedbrook2004} and experimental protocols with coherent states of light \cite{Grosshans2003,Lodewyck2007,Fossier2009,Jouguet2013,Huang2016,Zhang2020}. With nonclassical squeezed and entangled states, the continuous-variable (CV) protocols {\cite{Cerf2001,Garcia2009,Su2009}} become more robust and potentially applicable at distances up to 300 km \cite{Madsen2012} in optical fibers with attenuation 0.2 dB per kilometer, and tolerant to data processing inefficiency \cite{Usenko2011}. Such protocols can be advantageously implemented in both optical fiber links \cite{Madsen2012} and free-space atmospheric channels with realistic turbulence \cite{Derkach2018}. {Moreover, higher security can be offered  by relaxing the assumption about trusted devices for both coherent-state and entanglement-based protocols, as it was demonstrated by implementation of one-sided device-independent protocols \cite{Gehring2015,Walk2016}.}

A rate of secret key can be increased in CV QKD by frequency multiplexing of transmission channels \cite{Qu2017}. Frequency (wavelength) multiplexing is a well-known technique from classical optical communications \cite{Ishio1984}, also with the homodyne detection \cite{Grobe2013}. It can be similarly considered to increase the secret key rate of CV QKD protocols and has recently been studied using Gaussian modulation of  frequency comb states \cite{Wang2019},  as well as using independent lasers for each mode for subsequent  discrete \cite{Qu2017} or Gaussian modulation \cite{Kumar2019}. In this paper we test the entanglement-based CV QKD protocol that uses multiple frequency modes to multiplex the signal.
% However, crosstalk between different multiplexed modes, used for transmission, can be very destructive for CV QKD \cite{Li2018,Eriksson2019}, as a signal from one mode becomes a noise for the other modes.  
%Both modulation-based and entanglement-based frequency multiplexed CV QKD protocols suffer from unwanted cross-correlations (crosstalk) between the frequency modes that can be very destructive for security of the protocol. In modulation-based protocols the main source or crosstalk is the signal leakage between modes during fiber transmission \cite{Li2018,Eriksson2019, Wang2019}. The protocol we test in this paper is based on the bipartite entanglement between pairs of modes in the frequency comb. 
However, in practice
generated multiple entangled modes can become mutually coupled to each
other, resulting in cross-correlations as well as excess noise in the
modes, that can be destructive for CV QKD.  Importantly, homodyne detection of field quadratures gives sufficient information about states of light in the individual modes in order to compensate for the crosstalk. Based on these advantages, a crosstalk elimination based on {state or} data manipulations has been {addressed} in \cite{Filip2005, Usenko2018} demonstrating that such a limiting factor for multiplexed QKD can be {in principle} deterministically eliminated by optimized data manipulation, using the whole multimode structure (contrary to modes selection e.g. used to improve quantum steering in \cite{Cai2020}). Differently to protocols with single-photon detectors, it is therefore not required to implement an active strategy of optical decoupling which is very challenging for a large number of the transmitted modes.

Nowadays, CV QKD reaches a new level at which substantial increase of the secret key rate of mid-range protocols is a relevant target of the ongoing development {and requires, in particular, development of crosstalk elimination methods. The previously suggested methods either required heterodyne detection with optimal engineering of auxiliary input states and were not applied to CV QKD security analysis \cite{Filip2005} or considered only cross-talk interaction between the neighbouring modes of the otherwise perfect entangled states \cite{Usenko2018}. In the current paper we suggest the multimode crosstalk compensation method based on data manipulation, equivalent to linear state manipulations, experimentally test it on the real multiplexed entangled states {measured with the mode-discriminating homodyne detection suitable for CV QKD}, and evaluate the security of the resulting CV QKD protocol.}  Without such a test {on the multimode state with real crosstalk and errors} it is impossible to estimate applicability of {the crosstalk compensation} for a large number of modes. A positive result demonstrating that the secret key rate can be enhanced by channel multiplexing with high efficiency, despite crosstalk substantially reducing the achievable key rate already in the source, is necessary to open a pathway for further implementations and applications of frequency-multiplexed CV QKD. For the test we use simultaneously frequency multiplexed source of entanglement with 8 pairs of modes and {mode-discriminating} homodyne detection. For them, the crosstalk is a very natural phenomena pronouncedly reducing the key rate to a tiny number of 0.015 bits. We suggest and apply optimized data manipulation, which allows  decoupling of modes under crosstalk and brings large improvement to the achievable secret key rate. The secret key rate can be enhanced by almost a factor of 15. Reducing the channel noise in all frequency multiplexed channels by this data manipulation, alternatively, can extend the secure distance (channel range at which generation of secret key is still possible) by approximately 100 kilometres. Moreover, as the source emits femtosecond pulses, allowing for high system clock rates, the performance of the system can be also further increased by time multiplexing. Our result solves the major problem of mode crosstalk in the source, however, it can be equally applicable to crosstalk in the link and detection {(although the mode coupling inside an optical fiber is weak, if present at all, hence one can expect the mode interaction in the source to be the dominating cause of crosstalk).} Therefore, it opens the possibility for high-speed and high-capacity entanglement-based CV QKD with femtosecond frequency-multiplexed states.

\section{Results}
%We address the prerequisites for CV QKD using frequency multiplexed entangled states, deteriorated by crosstalk between frequency modes. By means of frequency-sensitive homodyne detection, we show the possibility to apply optimized local data manipulations in order to improve the frequency multiplexed resource and make the CV QKD more efficient and robust.

%
\begin{figure}[b]
\centering \includegraphics[width=0.9\linewidth]{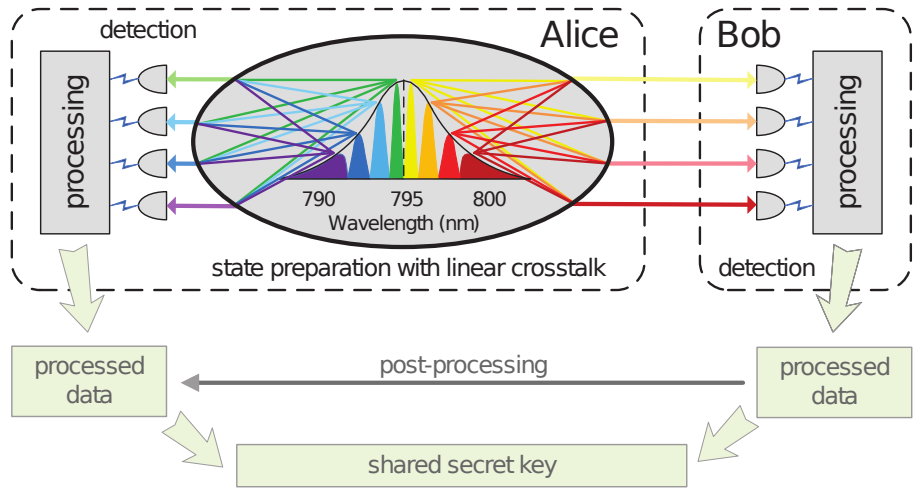}
\caption{Bright colors show a sketch of a CV QKD test-bed for study of the multimode entangled source at the side of sender, Alice, with crosstalk coupling between the frequency modes in both of the two beams, leaving the source. The entangled source is based on eight pairs of modes, where only four of them are shown for clarity. {We consider a scenario where half} of the modes (below the central frequency) is locally measured by Alice and another half (above the central frequency) transmitted to a remote trusted party Bob (trusted devices are given in dashed blocks). Both multimode beams are detected by homodyne detectors and processed to optimally eliminate the crosstalk and improve the secret key rate.  The data processing corresponds to a local physical multimode symplectic transformation and was optimized to achieve higher key rate between the trusted parties. The trusted parties then can use authenticated classical channel to perform post-processing by correcting their errors and amplifying the data privacy in order to obtain quantum-secure key as the result (this part of the protocol was modelled numerically so is illustrated in pale colors).}
\label{qkd}
\end{figure} 

We consider the use of entanglement source in multimode CV QKD test-bed based on the frequency multiplexed femtosecond pulses of light, consisting of 16 modes and with mode-discriminating homodyne detection, as described in Fig. \ref{qkd}. In our proof-of-principle experiment, all the 16 modes are generated in a single beam, and to test the applicability of the source for QKD purpose, we assume that the lower half of the frequency modes are distributed to Alice, and the other is to Bob.

\begin{figure}[t]
\centering \includegraphics[width=0.9\linewidth]{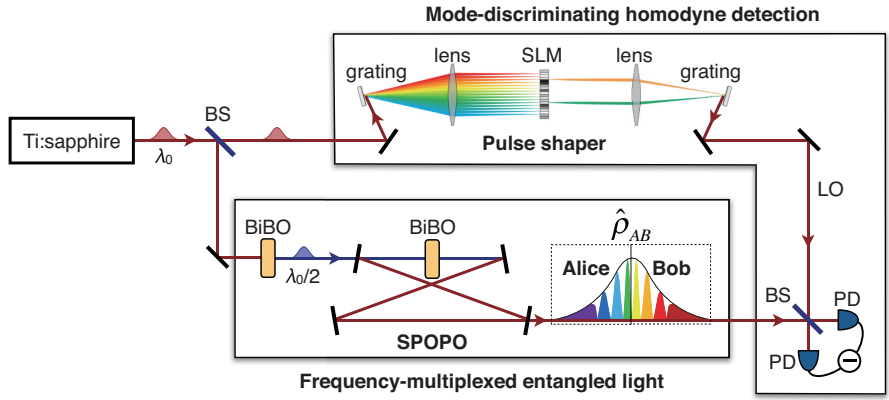}
\caption{Experimental setup for generation of frequency-multiplexed multimode entangled light and its measurement with mode-discriminating homodyne detection. The generated multimode light $\hat{\rho}_{AB}$ is in 16 frequency modes, where Alice (A) and Bob (B) access to the eight lower frequency modes and the other eight frequency modes, respectively. The pulse shaper is constructed in the folded configuration in actual implementation. BS: beam splitter; SLM: spatial light modulator; PD: photodiode. See main text for details.}
\label{setup}
\end{figure}

The experimental setup is shown in Fig. \ref{setup}. The main laser is a Ti-sapphire pulse laser, having a duration of 120 fs centered at $\lambda_0$ ( = 795 nm) with a repetition rate of 76 MHz. The beam from the laser splits into two beams, where one is used for generating frequency-multiplexed entangled light, and the other serves as a local oscillator (LO) for mode-discriminating homodyne detection. To generate the entangled light, we employ a synchronously pumped optical parametric oscillator (SPOPO) including a 2-mm-thick BiB$_3$O$_6$ (BiBO) crystal, which operates below the threshold \cite{Medeiros2014, deValcarcel2006}. The pump laser for the SPOPO (centered at $\lambda_0 / 2$) is prepared by second-harmonic generation of the main laser in a 0.2-mm-thick BiBO crystal. As a result, an entangled state of femtosecond pulses of light in multiple frequency modes (centered at $\lambda_0$) is generated in a single beam, and the efficiency of the process is enhanced by the cavity constituting the SPOPO. For our purpose, we consider sixteen frequency-band modes of the generated {multimode} light, and assume that the lower half (eight) frequency modes are measured by the trusted sender Alice, while the other half frequency modes are measured by the trusted receiver Bob after a multimode channel. We stress that even if in practice this separation is not performed in the current experiment, spectrally splitting a beam in two halves  {can be} readily done experimentally with a simple dispersive element, such as a grating, a dichroic mirror, or a prism. { It is possible to use a high efficiency grating or prism, or fiber based wavelength division multiplexing \cite{Liu2016}. These dispersive elements would introduce only small additional losses, leading to the excess noise in the generated multimode states. Such noise can be however considered trusted and will only have a limited negative effect on the key distribution \cite{Usenko2016}.}

To measure the generated {multimode} state, we use homodyne detection which can discriminate different frequency modes. {As the LO of homodyne detection determines the frequency mode, we control the LO based on a pulse shaping technique. For this purpose, a pulse shaper in the 4-f configuration is employed: an input beam is diffracted by an optical grating (1200 grooves/mm), which is subsequently focused by a cylindrical lens (190-mm focal length). On the Fourier plane of the lens, a reflection-type spatial light modulator having 512 x 512 pixels controls the amplitude and the phase of frequency modes. The reflected beam comes back to the lens and the grating. The overall wavelength resolution is found to be 0.1 nm.} Using the pulse shaper, a covariance matrix associated with the sixteen frequency modes was obtained by measuring quadrature outcomes in a sequential way from the {mode-discriminating} homodyne detection; in the homodyne detection, the two photodiodes have a quantum efficiency of 99 \%, fringe visibility is 93--95 \%, and demodulation frequency is 1 MHz \cite{Cai2017}. {The obtained covariance matrix is presented in Fig. \ref{cov_matrices} of Appendix A.}

Given the resolution of the pulse shaper, we consider that all the measured modes are realistically matched to the local oscillator. It also does not limit the applicability of the method which can be applied to the crosstalk in the multimode detector equally well. If the unmatched modes are present, they will contribute to noise \cite{Usenko2015a} and may act as a detection side channel \cite{Derkach2016}, but can be compensated for by increase of the brightness of the local oscillator \cite{Kovalenko2019}.

{To extend and verify the method of decoupling following the preliminary theoretical studies} \cite{Filip2005, Usenko2018} for the source depicted in Fig.\ref{qkd}, we assume a typical QKD scenario, where we suppose that Alice's preparation is trusted (being fully out of control by an eavesdropper Eve) and Alice is measuring her modes locally by a multimode homodyne, while the Bob's modes travel directly towards his detection. Bob is measuring his modes using mode-discriminating
homodyne detectors, also assumed to be trusted (including the efficiency and the electronic noise of the detectors). Ability to address the individual local modes in the homodyne detection is crucial for channel multiplexing in CV QKD and the multimode structure of entangled states can be harmful for the protocols otherwise \cite{Usenko2014}. To controllably investigate impact of lossy channel, we applied attenuation to Bob's measured results. It emulates an untrusted channel, characterized by the transmittance $T$, which is assumed to be fully controlled by an eavesdropper Eve, capable of collective attacks. We assume purely lossy (attenuating) channel as the background noise is already very small in real optical fiber channels, such approach allows modelling fiber as well as free-space channels, where fluctuations due to atmospheric turbulence are typically slow compared to the signal repetition rate \cite{Usenko2012}. 
 To comply with the experimental test-bed, where the multimode source was characterized, we assume that the crosstalk appears in the source, but our methodology can also be directly applied also to crosstalk in the channel and detectors.  

Security of CV QKD is evaluated as the positivity of the lower bound on the key rate, which, in case of collective attacks and reverse reconciliation \cite{Grosshans2002}, reads
\begin{equation}
K=max\{0,\beta I_{AB}-\chi_{BE}\},
\label{kreq}
\end{equation}
where $\beta \leq 1$ is the post-processing efficiency (further we realistically take $\beta=96\%$, which complies with the achieved post-processing efficiency \cite{Jouguet2011}), $I_{AB}$ is the mutual classical (Shannon) information between Alice and Bob, and $\chi_{BE}$ is the Holevo bound, which upper limits the information accessible to a potential eavesdropper Eve on Bob's measurement results. We address security against collective attacks for the Gaussian entanglement-based CV QKD \cite{Garcia2006, Wolf2006}, which can be directly extended to the finite-size regime \cite{Leverrier2010, Leverrier2010a} and implies security against general attacks \cite{Renner2009, Leverrier2013}. The reverse reconciliation is used to test secret key distribution for mid-range distance with channel attenuation below -3dB. The positivity of the lower bound (\ref{kreq}) implies that the trusted parties are able to distill the secret key with at least the rate $K$ by using classical post-processing (error correction and privacy amplification) \cite{Devetak2005}. We therefore analyze security of frequency-multiplexed CV QKD by evaluating the lower bound on the key rate per multimode channel use $K$ (further also referred to as the key rate). We follow the Gaussian security proofs and respective security analysis methods, as described in the Appendix A.
%We consider the experimentally generated and characterized multimode states for CV QKD over lossy quantum channels, assuming the same channel transmittance for all the modes, and evaluate the lower bound (\ref{kreq}) depending on channel transmittance, hence assessing the bounds on CV QKD realization using the multimode states. 

\begin{figure}[htb]
\minipage{0.65\linewidth}
\centering\includegraphics[width=0.99\linewidth]{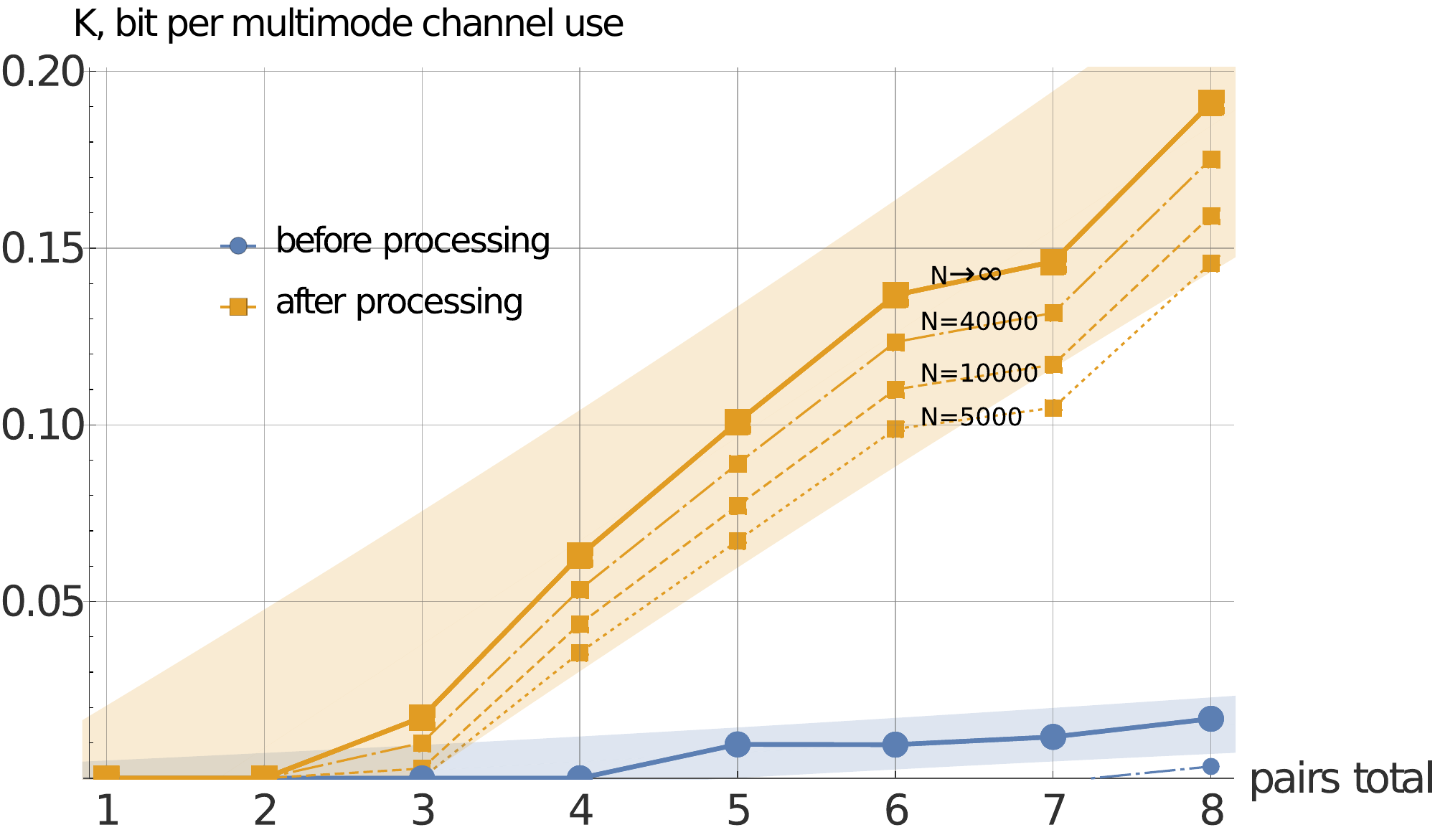}
\endminipage
\hfill
\minipage{0.35\linewidth}
\centering\includegraphics[width=0.99\linewidth]{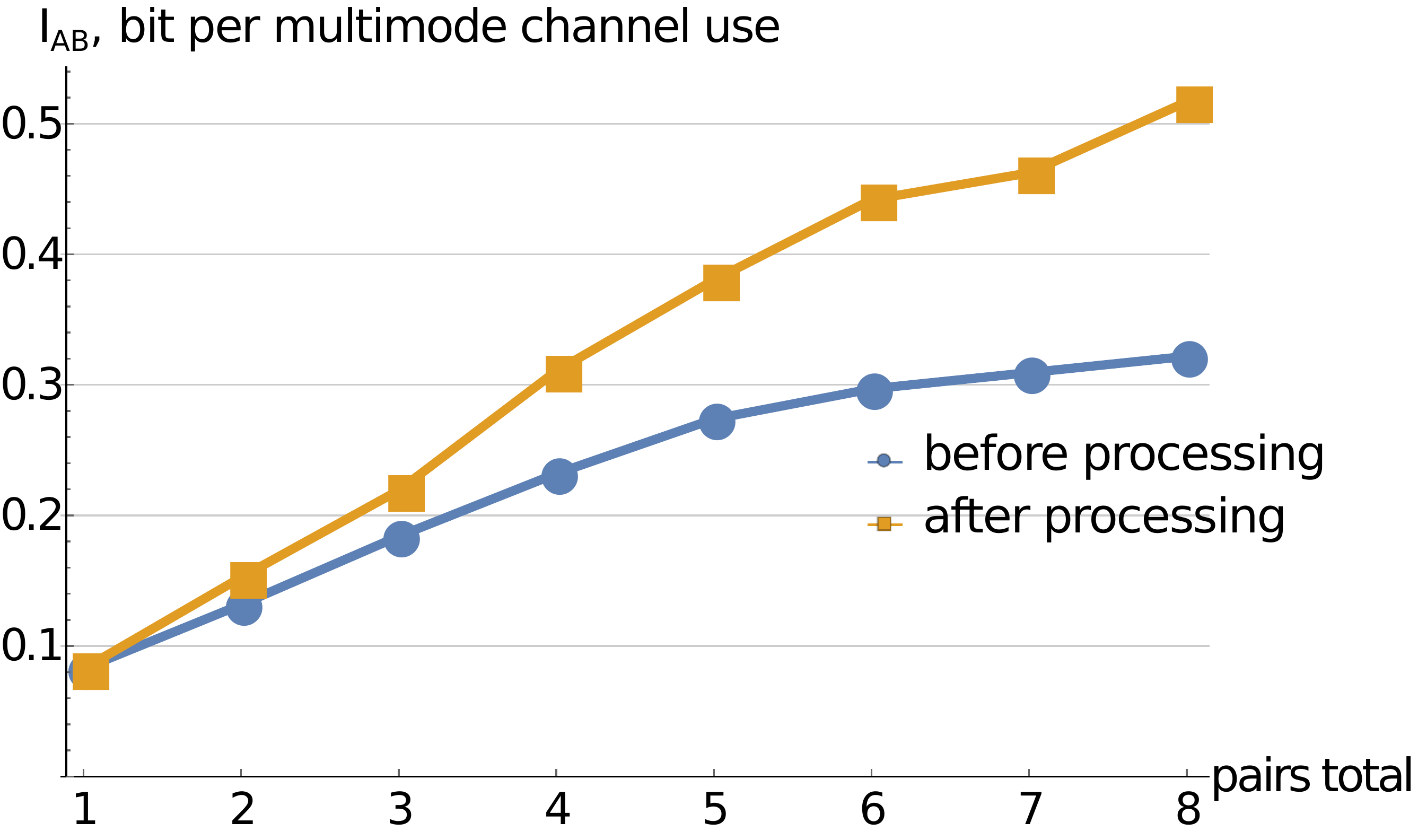} \\
\centering\includegraphics[width=0.99\linewidth]{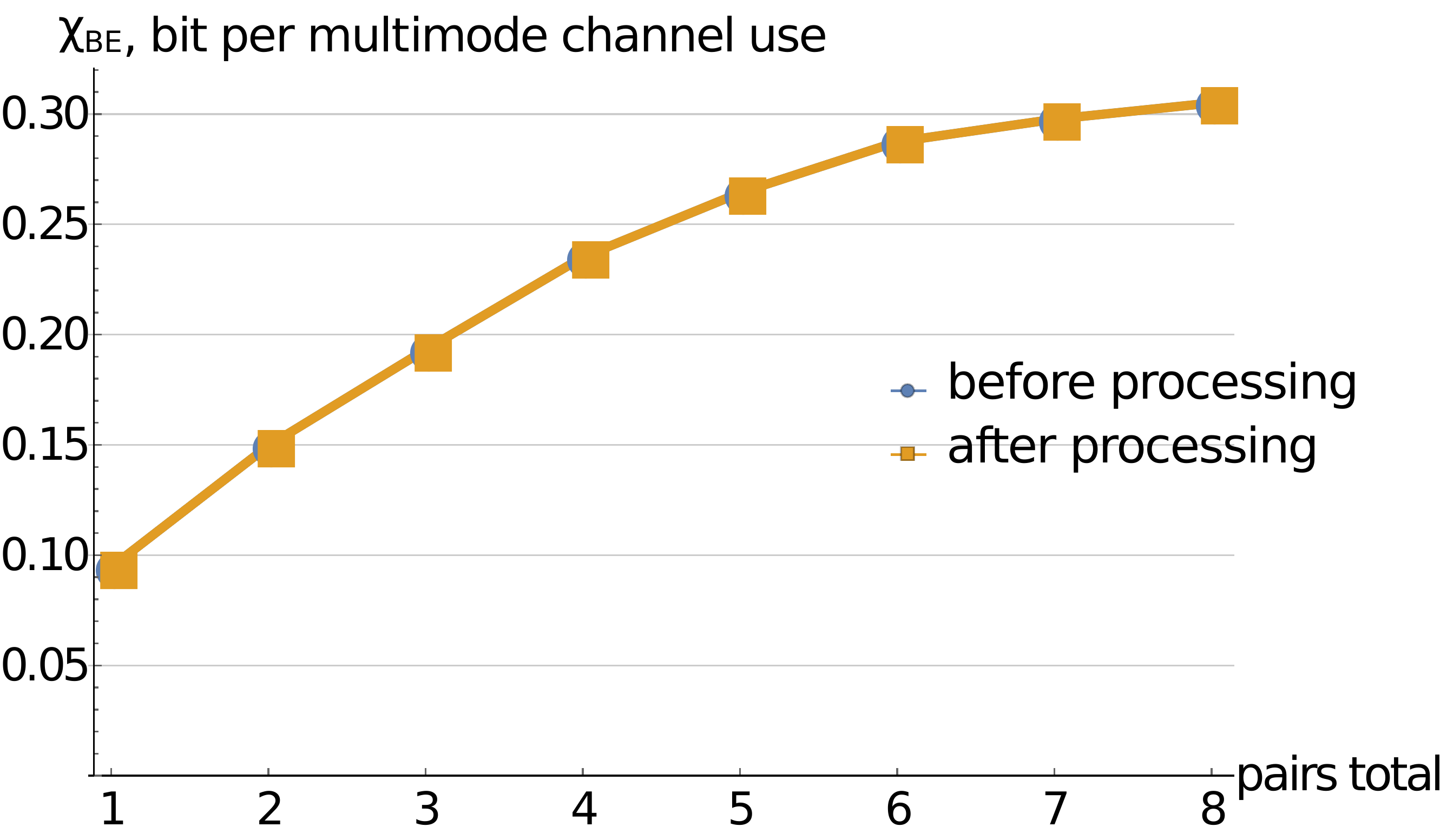}
\endminipage
\caption{Left panel: Estimated key rate (in terms of bits per multimode channel use) of CV QKD based on the frequency multiplexed entangled source in Fig.\ref{qkd} for different number of pairs measured by Alice
and Bob as obtained from the original data before decoupling (circles, blue lines) and after decoupling of modes involved in the multimode crosstalk by optimized data processing performed by the
trusted parties after the homodyne measurement (squares, yellow lines). The dotted, dot-dashed and dashed lines, represent the pessimistic estimates that take into account standard error for the respective number of measurements $N=5\cdot 10^3, N=10^4,N=4\cdot 10^4$, as indicated over the lines in the plots (see Appendix A for details). {Note that blue non-solid lines are almost extinct on the plot.} Shaded areas represent the method prediction bands with 95\% confidence level in the asymptotic limit of infinitely many measurement points. Realistic reconciliation efficiency $\beta=96\%$ taken for the processed data, perfect $\beta=1$ taken for the original data. While multiplexing brings only small and fragile advantage when all the pairs are being used, it can be drastically improved by optimized local data manipulations, revealing power of frequency multiplexing in CV QKD. The improvement gets more pronounced as the number of data points N increases. Right panel: Multimode mutual information (top) and Holevo bound (bottom) for different number of pairs measured by Alice and Bob as obtained from the original data (circles, blue line) and after optimized linear interactions performed by the trusted parties prior to the measurement (squares, yellow line). The plots illustrate the nature of improvement of frequency-multiplexed CV QKD by optimized data manipulations, which is based on increase of the mutual information, while the Holevo bound remains unchanged and, therefore, the yellow and blue points overlap.}
\label{kr-pairs}
\end{figure}

We demonstrate the power of the multimode states in CV QKD by confirming the gradual increase of the overall key rate for increasing the number of pairs of modes measured by Alice and Bob, as shown in Fig. \ref{kr-pairs} (left), assuming channel transmittance $T=0.2$. We first rank the pairs by the key rate between the individual pairs and then add pair by pair, thereby obtaining larger key rate, as seen in Fig. \ref{kr-pairs} (left, circles and the blue solid line). 
 
For the multimode states as shown in Fig.\ref{qkd}, we optimize the data processing { to achieve the maximum key rate (\ref{kreq}). The applied data processing is equivalent to local passive symplectic transformations when both the sender and the receiver separately act on their respective modes by optimized beam splitter networks} \cite{Weedbrook2012a}, see Appendix A for details. The results of the optimization are given in Fig. \ref{kr-pairs} (squares and solid yellow line) and it is evident, that optimized local data manipulations lead to almost 15-fold increase of the overall key rate for the multimode states. {The optimization process can, therefore, efficiently retrieve multiple pairs of entanglement from the multimode entanglement resource, providing significant improvement for CV QKD.} Moreover, the key rate becomes much more robust against statistical error in the finite data ensemble, as can be seen from the respective plots in Fig. \ref{kr-pairs} (squares and dashed yellow lines).

The improvement of multimode CV QKD by the local data processing is concerned with the increase of the multimode mutual information, as can be seen from Fig. \ref{kr-pairs} (right, top), while Holevo bound is not affected, as seen from Fig. \ref{kr-pairs} (right, bottom). Indeed, the local data processing (equivalent to symplectic transformations) does not affect the quantum entropies contributing to the Holevo bound by not changing the symplectic spectrum (i.e., thermal-state decomposition) of the multimode Gaussian state. On the other hand, redistribution of modes occupation by the local symplectic transformations increases the additive classical mutual information due to increased correlations, and can be optimized to achieve the best performance. This also substantially simplifies the optimization of the method by reaching the maximum mutual information, leading to the maximum key rate. Our method is focused on maximizing mutual information (not on eliminating cross-correlations) and leads to optimal improvement of the key rate. Note, that our method can be advantageously combined with the protocol, in which the Holevo information, maximally accessible to Eve, is minimized \cite{Jacobsen2018}. In this scenario by proper multiplexing and elimination of crosstalk higher key rate can be achieved at low post-processing efficiencies, while not increasing the information leakage. This can be particularly promising for high-speed CV QKD, where fast, but less efficient error correction can be otherwise a very limiting bottleneck \cite{Lodewyck2007,Usenko2011}.
\section{Discussion}
\begin{figure}[htb]
\centering \includegraphics[width=0.8\linewidth]{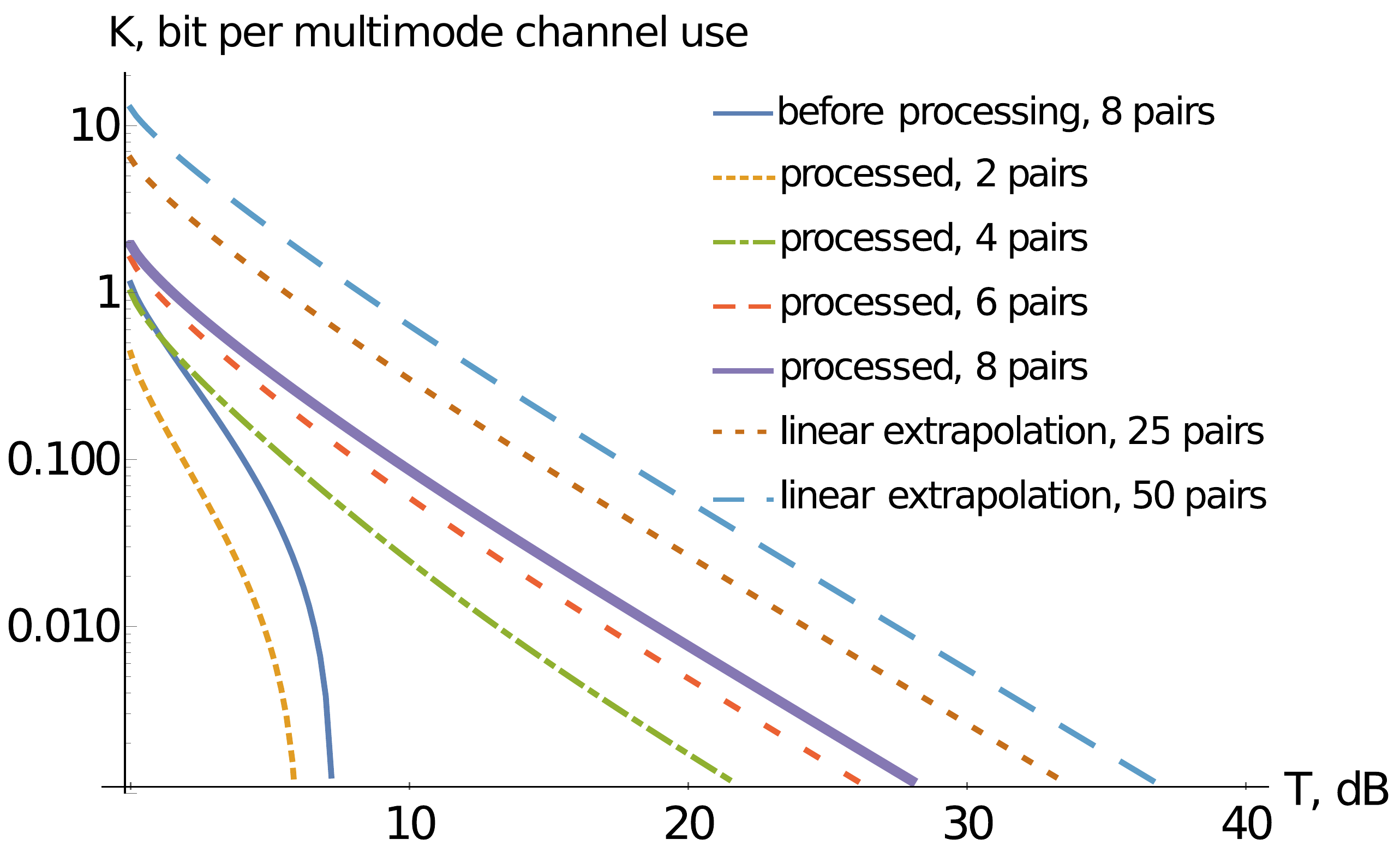}
\caption{Key rate of CV QKD versus channel transmittance T (in dB) as obtained from the original data on the full multimode entangled state (blue solid line), after optimized local data manipulations performed by the trusted parties for different number of used pairs of modes (non-solid lines for reduced number of pairs and thick solid violet line for the maximum number of eight pairs), linear extrapolation for larger number of modes (blue and brown dashed lines). Post-processing efficiency $\beta=96\%$. Evidently, optimized data manipulation can drastically improve robustness to loss (and, respectively, the secure distance) of frequency-multiplexed CV QKD with entangled states.}
\label{kr-pairs-dist}
\end{figure}

We also analyse the robustness of CV QKD against channel attenuation (which is equivalent to channel distance) with the full set of eight pairs of modes before and after the optimized data manipulation as shown in Fig. \ref{kr-pairs-dist} (the 8-pair covariance matrix after optimized data processing is illustrated in Fig. 4 in the Appendix A). By eliminating the crosstalk, we reduce the state preparation noise in the individual channels and respectively increase the maximum tolerable channel noise. It is evident from the plot, that optimized local data manipulation increases the maximum tolerable channel attenuation of the protocol from approx. 8 dB to {28} dB of loss, thus demonstrating potentially more than threefold increase of the secure distance of the multimode CV QKD protocol {(assuming 0.2 dB/km loss)}. We extrapolate the obtained results for the cases of 25 and 50 modes, which are expected to further improve the efficiency and robustness of CV QKD protocol up to approx. 34 and 36 dB, see Appendix A for details.
We also address the efficiency of our method by comparing the achieved results to the bounds set by eight times maximum performance of one best pair of modes (as the total number of pairs in our experiment is eight). In the way similar to maximization of the total key rate, we now run optimization to have as much as possible key in this particular pair. In terms of mutual information, the maximum in one pair  is 0.28 bit per channel use, the total maximum mutual information achieved by our method is 0.517 bit per channel use, the bound (eight times the maximum value for one pair) is 2.24 bit per channel use, which is 4.3 times larger than we achieve. For the key rate the maximum for one pair is 0.163 bit per channel use, the total achieved key rate is 0.212 bit per channel use, the bound is 1.304 bit per channel use, which is 6.15 time larger than we achieve. We define the decoupling efficiency as the ratio between the secret key rate achieved and the secret key rate that could be achieved in a perfect setting with all 8 pairs having maximal mutual Shannon information. The efficiency of our method for the secret key rate therefore reaches  0.16. This removable limitation is caused by source imperfections beyond the linear crosstalk and would require development of additional advanced experimental and data processing methods to further improve practical frequency-multiplexed CV QKD. Our results show that despite drastic improvement achieved with the suggested method for crosstalk elimination, even higher performance can be achieved by larger number of frequency channels with faster data processing and by further developed experimental techniques aimed at reduction of crosstalk.

\subsection*{Summary and outlook}
By optimally applying data manipulations we were able to compensate the crosstalk in the frequency multiplexed CV QKD with femtosecond-pulsed entangled states and substantially increase the mutual information between the sets of modes, measured by the trusted parties, while the leaked information, upper bounded by a function of Gaussian quantum entropies, did not change. Thus we can increase the achievable key rate for continuous-variable quantum key distribution or, equivalently, extend the secure distance of the protocols. The results of the optimized local data manipulations show the possibility to increase the overall key rate by almost the factor of 15 and extend the secure distance for the multiplexed entanglement-based protocol by the factor of three. Note, that while higher key rates can be as well obtained by increasing the system repetition rate, our method does not affect the information leakage and only increases the mutual information, hence drastically increasing the key. Nevertheless, our method can be further combined with the increase of the repetition rate to achieve even higher key rates. In the present demonstration, the calculations were performed on a covariance matrix obtained by mode-discriminating homodyne measurement and can be easily extended to large number of modes \cite{Thiel:2017es}. Furthermore, the crosstalk between the modes can also be further reduced or adapted to the measurement system manipulating the source through spectral shaping of the pump \cite{Arzani:2018fq}.  While we applied data manipulations to compensate crosstalk in the multimode CV QKD source, the method can be also used to eliminate crosstalk that appears in the multimode quantum channel. Our method is therefore very promising for improving key rates of continuous-variable quantum key distribution and can also be combined with the protocol based on minimization of the information leakage \cite{Jacobsen2018}, especially with elimination of channel noise and efficient channel estimation techniques, in order to overcome the limitations imposed by realistic fast post-processing. Moreover, we can combine our method with the existing tools to eliminate correlated noise \cite{Lassen2013} and side channels \cite{Derkach2016}. We therefore open the pathway to very high-speed practical realization of quantum key distribution using continuous variables. It should be followed by a test of complete multiplexed protocol together with secret key generation and can be extended to networking entanglement-based communication settings. Furthermore, the suggested crosstalk compensation technique can be useful in other applications of continuous-variable quantum information, such as quantum imaging \cite{Genovese2016} or quantum illumination \cite{Lopaeva2013}.

\section*{Appendix A}

\subsubsection*{Security analysis}

The key rate is calculated as $K= \max \left\lbrace  0,\beta I_{p_{AB}}-\chi_{BE} \right\rbrace  $, where
$I_{p_{AB}}$ is the classical information between Alice and Bob in $\hat{p}=i(\hat{a}^\dagger-\hat{a})$ quadrature (We chose it because in this experiment it gives larger key than the $\hat{x}=\hat{a}^\dagger+\hat{a}$ quadrature).
%Eavesdropper's information is given by the Holevo bound $\chi_{BE}$, i.e. we assume that Eve performs collective measurement on all 8 modes she has access to. The Shannon information between Alice and Bob is calculated as a sum of information in each pair $I_{p_{AB}}=\sum_{i=1}^8{I_{p_{A_iB_i}}}$, i.e. we take into account only pair by pair correlations and disregard all other correlations. These other correlations are treated as cross talk or extra noise that increases Eve's information but does not contribute to the key rate. Reducing correlation among the pairs and increasing correlations inside the pairs we can increase the Shannon information and the secret key rate. To do so we can introduce any local operations on either side (sender or receiver).

The classical mutual information for a pair of Gaussian-distributed data sets A and B with variances $V_A$ and $V_B$ respectively can be evaluated as $I_{AB}=\log_2{(V_A/V_{A|B})}$, where $V_{A|B}=V_A-C_{AB}^2/V_B$ is the conditional variance, which can be expressed through the correlations between the data sets, $C_{AB}$. It is therefore straightforward to evaluate our multimode mutual information $I_{AB}=\sum_{i=1}^8{I_{A_iB_i}}$, which is the sum of bipartite mutual information quantities between eight pairs of data sets obtained from the homodyne measurements of different frequency modes on both Alice's and Bob's sides. Note that here and further the mutual information as well as the lower bound on the key rate (1) is evaluated in bits per {multimode} channel use. 

The calculation of the Holevo bound is more involved and is performed in the assumption that Eve is capable of collective measurement of the eight-mode state, reflected from the attenuating channel, similarly to the single-mode CV QKD in purely lossy channels \cite{Grosshans2005}, as Eve's vacuum modes, corresponding to the loss in each of the modes, cannot be correlated. The Holevo bound is then evaluated as the difference $S(E)-S(E|B)$ between the von Neumann (quantum) entropies of the state available to Eve prior and after conditioning on the measurements of the receiving trusted party Bob.  The von Neumann entropy of a state described by covariance matrix $\gamma $ is calculated  as $S(E)=\sum_i \text{G } \frac{\lambda_i-1}{2}$, where $\lambda_i$ are symplectic eigenvalues of $\gamma_E$ and $\text{G}(x)=(x+1)\log_2(x+1)-x\log_2 x$. Here $S(E)$ is the entropy of the eight-mode state measured by Eve, and $S(E|B)$ is the entropy of Eve's state, conditioned on the set of $\{x_{B_i}\}$, being the measurement outcomes of the homodyne detection in $x$-quadrature on eight modes at Bob's station (equivalently for the $p$-quadrature measurements). The calculation is performed in the covariance matrix formalism, within the pessimistic Gaussian state approximation (see more details on the Gaussian security analysis in \cite{Usenko2016}).

\subsubsection*{Error estimation}

To estimate the effects of the measurement error on the key rate, we assume that every pair of modes has bi-variate normal quadrature distributions and covariance matrix for the $i,j$-pair is
\begin{equation}\gamma_{ij}=
\left(
\begin{matrix}
V_i & C_{ij}\\
C_{ij}  &  V_j  \\
\end{matrix}
\right).
\label{covmat}
\end{equation}
Here $V_i = \left(
\begin{matrix}
\langle\Delta x_i^2\rangle & 0\\
0 &   \langle\Delta p_i^2\rangle \\
\end{matrix}
\right)$  and $C_{ij} = \left(
\begin{matrix}
\langle x_i x_j\rangle & 0\\
0 &   \langle p_i p_j\rangle \\
\end{matrix}
\right)$, as in this experiment no correlations between quadratures were observed, hence $\langle x_i p_j\rangle=0$ $\forall i,j$. The best estimate for the standard error for $\gamma_{i,j}$ after N measurements will be  \cite{Kendall1987}
\begin{equation}\sigma_{\gamma_{i,j}}= \frac{1}{\sqrt{N+1}}
\left(
\begin{matrix}
\sqrt{2} V_i & \sqrt{V_i V_j+C_{ij}^2}\\
\sqrt{V_i V_j+C_{ij}^2} & \sqrt{2}  V_j \\
\end{matrix}
\right).
\label{errormat}
\end{equation}
We assume the worst case (pessimistic) scenario for different numbers of measurements and evaluate the lower bound on the secret key rate for each case. The pessimistic scenario implies that the diagonal elements of the covariance matrix (variances) are increased by the error value while the absolute value of the off-diagonal elements (correlations) are decreased \cite{Ruppert2014,Leverrier2015}. The results are presented in the Fig. 2 as dashed lines. It is evident from the plots in the Fig. 2, that even multiplexing of all eight pairs can only slightly restore the non-zero key rate if the measurement results are used without any processing, and that the key rate is very sensitive to the error in the finite data samples. One can expect that the performance of the multiplexed CV QKD is strongly limited by the crosstalk between the modes \cite{Usenko2018}, which is likely to appear in the generation of frequency multiplexed entangled states under study. We therefore suggest and verify the method of optimized data manipulation after the homodyne detection, performed by the trusted sides, in order to substantially compensate the crosstalk and make the multimode resource more applicable for CV QKD. 

\subsubsection*{Optimized data processing}

Generally all the sixteen modes are getting coupled in the state preparation and such crosstalk should be possible to at least partially compensate for using a global 16x16 symplectic transformation that maximizes the mutual information. The quantum communication scenario {makes such global transformation impossible, but the crosstalk can be significantly reduced even when} we consider Alice and Bob performing only local operations independently of each other. {Both Alice and Bob each control 8 modes of the shared 16-mode state. Each of them can then introduce linear local passive operations on their respective sides in order to minimize the crosstalk while preserving the security of the protocol.} We are therefore looking for two  8x8 local symplectic transformation matrices equivalent to a sequence of linear optical devices.

{The covariance matrix of the whole 16-mode state $\gamma$ can be represented as 
\begin{equation}
\gamma = \left( {\begin{array}{ccc}
   V_1& \cdots& C_{1,16}\\
   \vdots&\ddots&\vdots\\
   C_{16,1}&\cdots & V_{16} \\
  \end{array} } \right), 
  \end{equation} with $V_i$ and $C_{i,j}$ given in eq. (2) and below. 
To model the interaction we assume that a set of $2\times 2$ beam splitters is introduced between all possible pairwise mode permutations on the same side.  i.e. there is $(N/2-1)N/2=56$ beam splitters \cite{Reck1994} (28 on Bob's and 28 on Alice's side).
The phase convention we use for a $2\times 2$ beam splitter acting on a 16-mode state is 
\begin{equation}
 T_{ij}=
  \left( {\begin{array}{cccccc}
  \mathbb{I} &\cdots&0&0&\cdots&0\\
  \vdots&\ddots&\vdots&\vdots& &\vdots\\
  0 &\cdots& \sqrt{t_{ij}}~\mathbb{I} & \sqrt{1-t_{ij}}~\mathbb{I}&\cdots&0 \\
  0 &\cdots& \sqrt{1-t_{ij}}~\mathbb{I} & -\sqrt{t_{ij}}~\mathbb{I}&\cdots&0 \\
     \vdots& &\vdots&\vdots&\ddots&\vdots\\
   0 &\cdots&0& 0 &\cdots&\mathbb{I}
  \end{array} } \right),
\end{equation}
where $t_{ij}$ is the transmittance coefficient, $i, j$ are the modes that are interacting on the given beam splitter.
Then introducing the beam-splitter network on the sender side is equivalent to Alice acting on the covariance matrix with the sequence of the beam splitter two-mode linear coupling operation: first Alice acts with $\gamma'=T_{1,2}\gamma T_{1,2}^T$, then $\gamma''=T_{1,3}\gamma' T_{1,3}^T$ etc. As a result the sender (Alice) transforms  the initial state with the product of 28 operators \begin{equation} 
U_{A}=T_{7,8}T_{6,8}...T_{1,3}T_{1,2} =\prod_{i=1,j=i+1}^8  T_{i,j}
\end{equation}
 on her side and the receiver (Bob) acts in the same manner with the operation 
\begin{equation}
 U_{B}=T_{15,16}T_{14,16}...T_{8,10}T_{8,9}=\prod_{i=9,j=i+1}^{16}  T_{i,j}
\end{equation}
 on his side. Their joint interaction operation is 
$U = U_A U_B$.}  After the beam-splitter network is applied to the original state, the covariance matrix becomes $\gamma_{f} = U \gamma U^T $.

\begin{figure}[htb]
\minipage{0.45\linewidth}
\centering\includegraphics[width=0.9\linewidth]{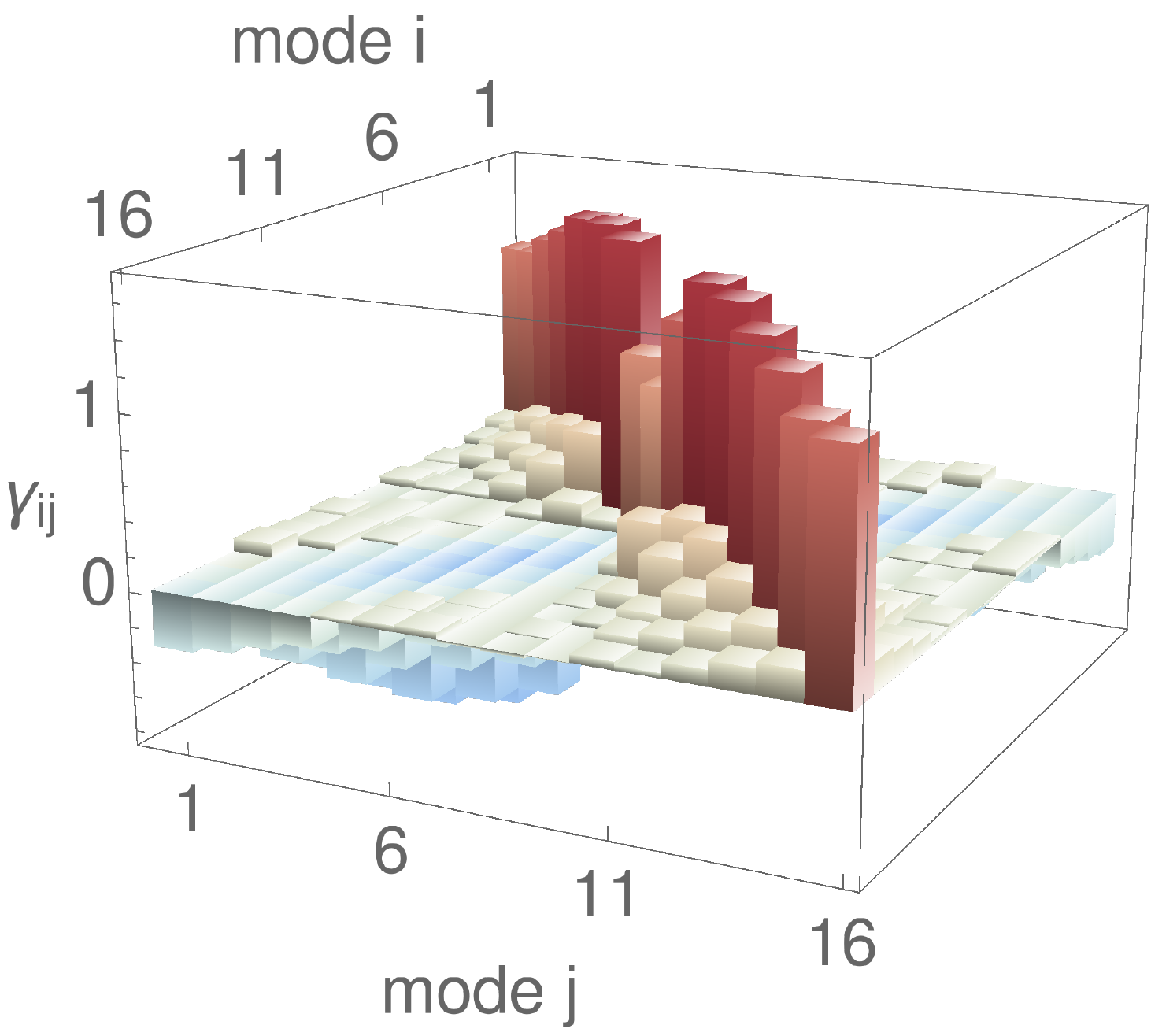}\\
\centering\includegraphics[width=0.9\linewidth]{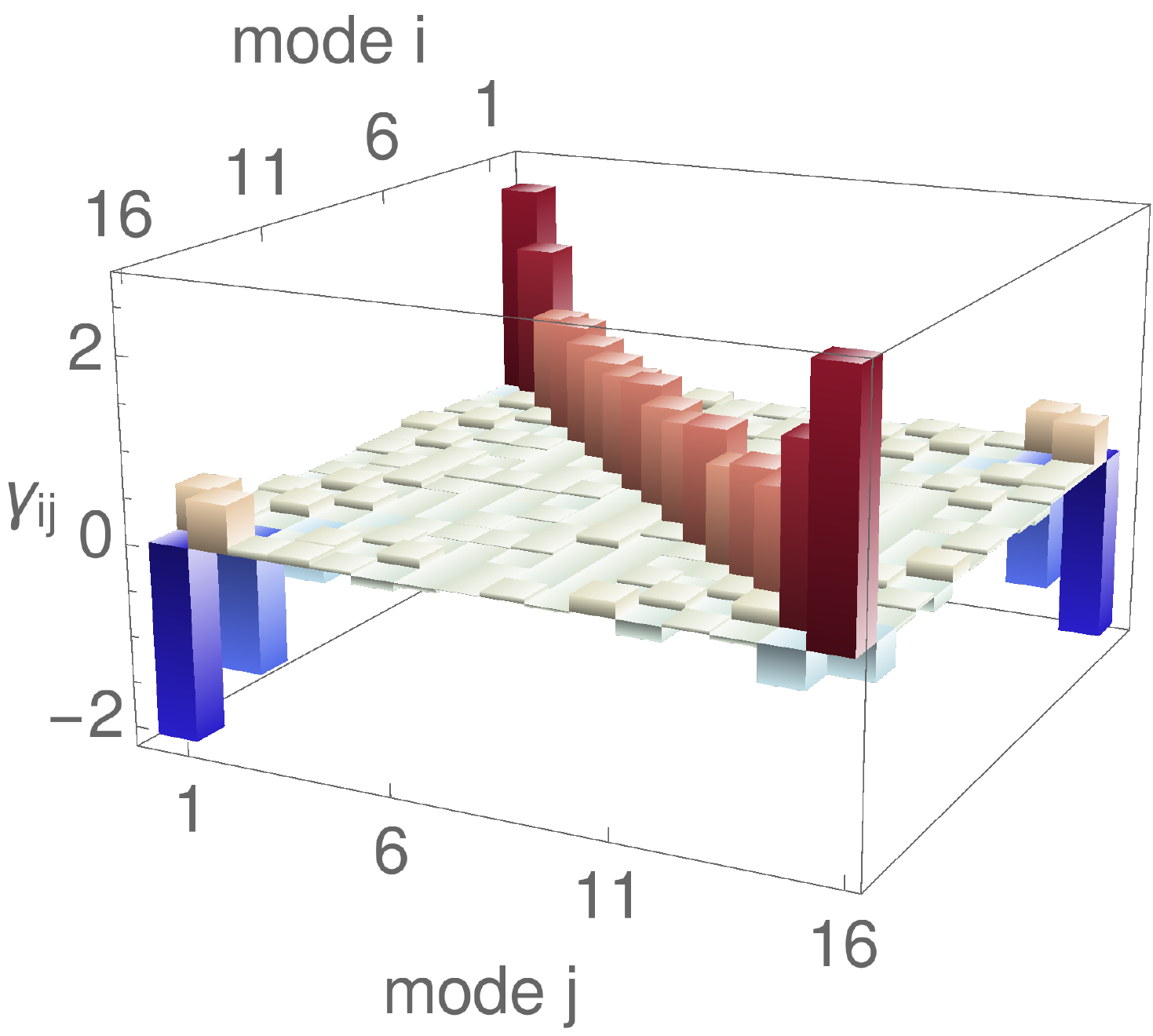}
\endminipage
\hfill
\minipage{0.45\linewidth}
\centering\includegraphics[width=0.9\linewidth]{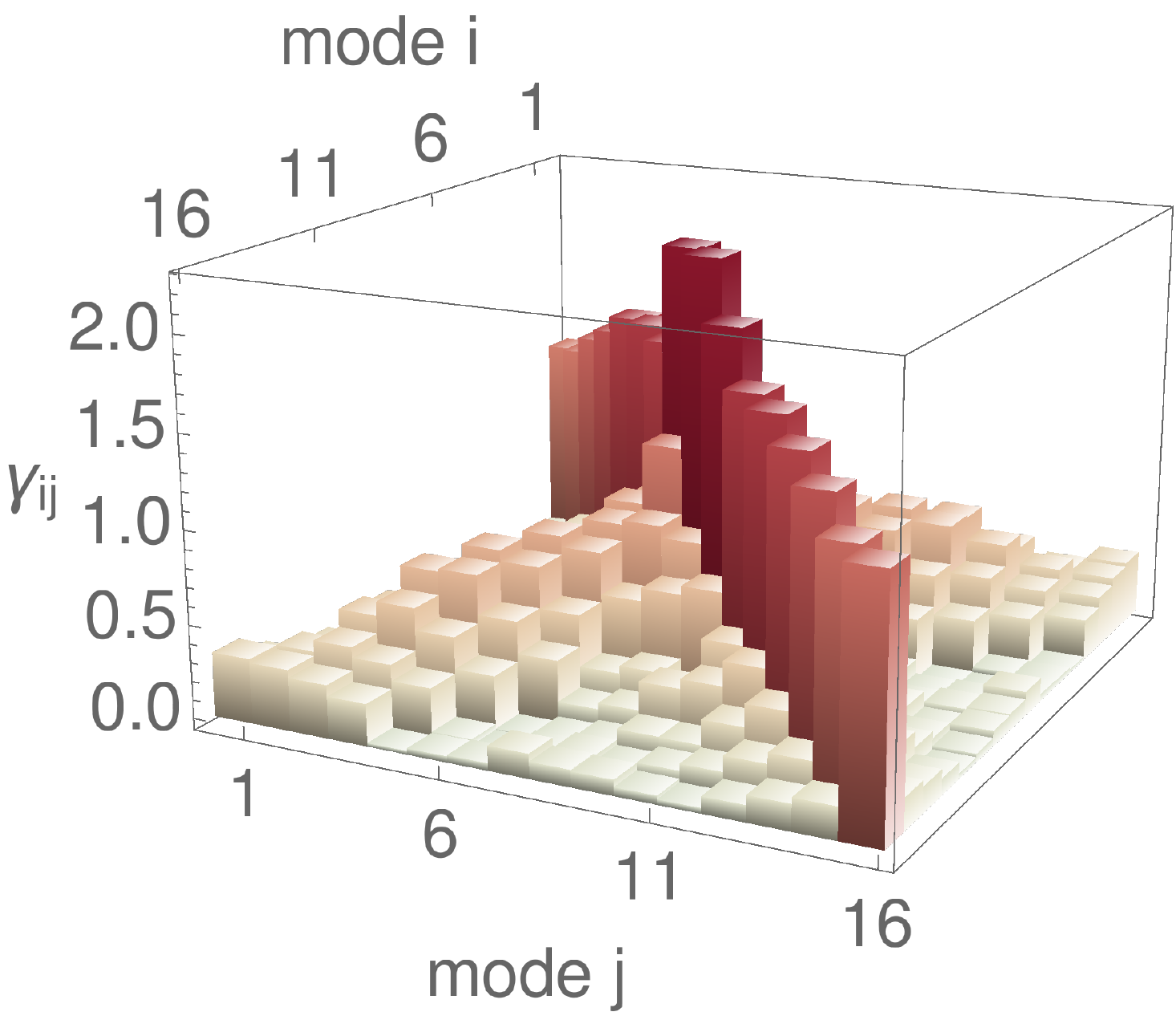} \\
\centering\includegraphics[width=0.9\linewidth]{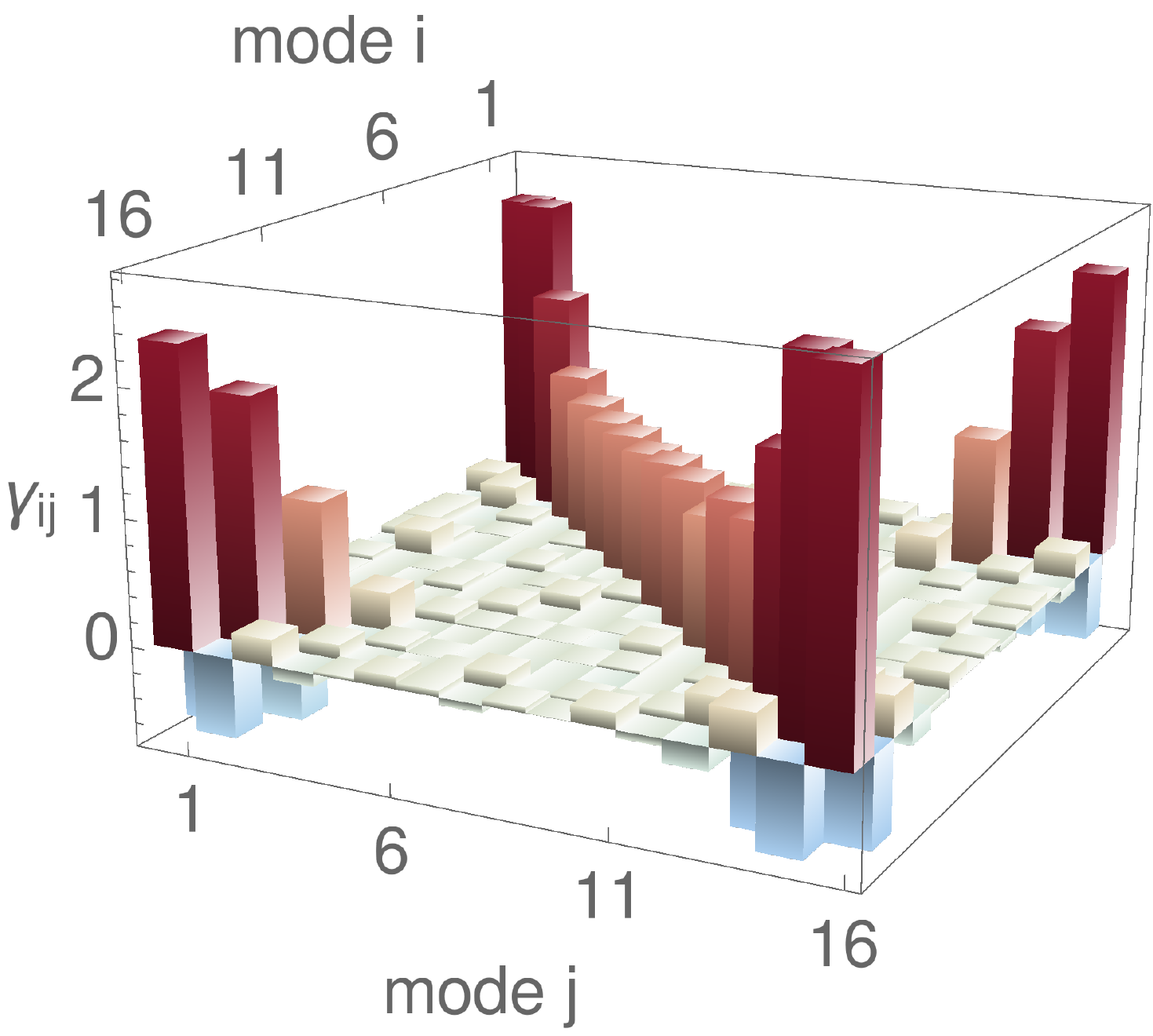}
\endminipage
\caption{Visualization of covariance matrices in $\hat{x}$ quadrature (left) and $\hat{p}$ quadrature (right) before (top) and after (bottom) the optimized linear data processing.}

\label{cov_matrices}
\end{figure}

We then calculate the mutual information { $I_{x_{AB}}$ and $I_{p_{AB}}$} of the state $\gamma_{f}$ separately in $\hat{x}$ and $\hat{p}$ quadratures and  {maximize} the functions $I_{x_{AB}} (\textbf{t})$ and $I_{p_{AB}} (\textbf{t})$ numerically, {here $\textbf{t}=(t_{1,2},t_{2,3}...t_{15,16})$ is the variable vector made of transmittance coefficients of the beam splitters.}
There is no need to maximize the key rate, as the Holevo bound isn't affected by unitary transformations (indeed, the von Neumann entropy of the states  {is} preserved, {hence the maximization of the mutual information is sufficient}). The optimization was done numerically using limited memory Broyden--Fletcher--Goldfarb--Shannon (l-BGFS) optimization algorithm with bound constraints \cite{Byrd1994} from SciPy library. {The l-BGFS performs  $O(d)$ computation per iteration, where $d$ is the number of the function's variables, in this case $d=(N/2-1)N/2$ and the method performance scales depending on the number of the modes as $O(N^2)$. In general,  l-BGFS does not converge to a global {maximum} if the function under {maximization} is not a {convex} one as is the case here}. To find the global {maximum} we used the basin-hopping optimization method. Naturally there is no guarantee that each {maximum} we have found is indeed a global one, but the obtained results already shown drastic improvement of quantum communication using the multimode states. The visualization of the covariance matrices before and after the optimization is given in Fig. \ref{cov_matrices}{, where the raw data used in the optimization are reported in \cite{Cai2017}.} It shows noticeable redistribution of correlations between the modes.

{It is worth mentioning that while here we optimize the state with only passive local operations, it is also possible to use an active transformation, although it would significantly increase the computing power needed. Besides}, the most general set of {passive} local operations would be represented by sequence of Mach-Zehnder interferometers with beam splitters of optimized transmittance and optimized phase shifts between them. We have checked this kind of optimization setup as well but it did not help to increase the Shannon information and the secret key rate. This is due the fact that the correlations between $\hat{x}$ and $\hat{p}$ quadratures in the data are negligibly small.

{The matrix of the optimized interaction is to be found on the parameter estimation step  of the QKD protocol {based on the estimation of the state, shared between the trusted parties, in terms of its covariance matrix \cite{Leverrier2015}}. The optimization can be performed on either sender or receiver side and then announced publicly. {Since the optimization parameters are not related to the raw key data, no further disclosure and discarding of the key bits is needed.} Eavesdropper's knowledge of the optimized interaction does not influence security of the protocol as the security proof already assumes eavesdropper's ability to perform an {optimal collective} measurement {on the intercepted signal \cite{Grosshans2005} and the Holevo bound is not affected by the linear interactions between the signal modes  on the trusted sides.}}

\subsubsection*{Results extrapolation}

We predict the efficiency of our method for larger number of pairs, by evaluating prediction bands, as seen in Fig. \ref{kr-pairs} (left). To do so, we first used a linear fit for the key rate results in order to predict how the key rate will behave if we add more modes. If the pairs of modes were uncorrelated (i.e., experience no crosstalk) and all had the same variance, the key would grow linearly, therefore we assume that in our case of correlated modes dependence will stay close to linear. Using the method of the least squares \cite{Bevington2003} we got a linear model for the key rate in the form $K(x)=a+b x$ (for the processed data we have $a=-0.0501$ and $b=0.0293$). We then evaluated the prediction bands defined as $K(x)\pm t \sqrt{s^2+X \mbox{Cov} X^T}$, where $\mbox{Cov}$ is the covariance matrix for the coefficients $a$ and $b$, and $s^2$ is the mean squared error for the data points,
$X = \left(
\begin{array}{c}
 1  \\
 x \\
\end{array}
\right)$,
$t$ is defined from the Students distribution for 95\% confidence level (resulting in $t=2.447$).

\section*{Funding}

O.K. acknowledges project IGA-PrF-2021-006 of Palacky University; Y.-S.R. acknowledges support from the European Commission through Marie Sk\l{}odowska-Curie actions (Grant No. 708201) and from a National Research Foundation of Korea grant funded by the Korea government Ministry of Science and ICT (Grant No. NRF-2019R1C1C1005196); Y.C acknowledges projects 11904279 and 12174302 of National Natural Science Foundation of China; O.K. and V.C.U. acknowledge project 19-23739S of the Czech Science Foundation; R.F. acknowledges project CZ.02.1.01/0.0/0.0/16\_026/0008460 of the Czech Ministry of Education and national funding from the MEYS and the funding from European Union's Horizon 2020 (2014-2020) research and innovation framework programme under grant agreement No 731473 (project 8C20002 ShoQC). Project ShoQC has received funding from the QuantERA ERA-NET Cofund in Quantum Technologies implemented within the European Union's Horizon 2020 Programme.
The authors also acknowledge support from COST Action CA 15220 QTSpace. The research leading to these results has received funding from the H2020 European
Programme under Grant Agreement 820466 CIVIQ and 951737 NONGAUSS.

\section*{Disclosures}

The authors declare no conflicts of interest.

\section*{Data availability}

{Data that support the findings of this paper are not publicly available at this time but may be obtained from the authors upon reasonable request.}

%%%%%%%%%%%%%%%%%%%%%%% References %%%%%%%%%%%%%%%%%%%%%%%%%

%%%%%%%%%% If using BibTeX:
\bibliography{combs_qkd}

%%%%%%%%%% If preparing manually:
% \begin{thebibliography}{1}
% \newcommand{\enquote}[1]{``#1''}

% \bibitem{Zhang:14}
% Y.~Zhang, S.~Qiao, L.~Sun, Q.~W. Shi, W.~Huang, L.~Li, and Z.~Yang,
%   \enquote{Photoinduced active terahertz metamaterials with nanostructured
%   vanadium dioxide film deposited by sol-gel method,}
%   {\protect\JournalTitle{Optics Express}} \textbf{22}, 11070--11078 (2014).

% \bibitem{OSA}
% {Optical Society}, \enquote{{OSA Publishing},}
%   \url{http://www.osapublishing.org}.

% \bibitem{FORSTER2007}
% P.~Forster, V.~Ramaswamy, P.~Artaxo, T.~Bernsten, R.~Betts, D.~Fahey,
%   J.~Haywood, J.~Lean, D.~Lowe, G.~Myhre, J.~Nganga, R.~Prinn, G.~Raga,
%   M.~Schulz, and R.~V. Dorland, \enquote{Changes in atmospheric consituents and
%   in radiative forcing,} in \enquote{Climate Change 2007: The Physical Science
%   Basis. Contribution of Working Group 1 to the Fourth assesment report of
%   Intergovernmental Panel on Climate Change,}  S.~Solomon, D.~Qin, M.~Manning,
%   Z.~Chen, M.~Marquis, K.~B. Averyt, M.~Tignor, and H.~L. Miler, eds.
%   (Cambridge University Press, 2007).

% \end{thebibliography}

\end{document}